# Crucial Experiment to resolve Abraham-Minkowski controversy

Zhong-Yue. Wang[1] , Pin-Yu Wang, Yan-Rong Xu

*Laboratory of Physics*

*Zhejiang Agriculture & Forestry University (ZAFU)*

*No.88 North Ring Road,* Lin'an, *Hangzhou, Zhejiang, 311300, China*

**Abstract:** Abraham-Minkowski dilemma concerning the momentum of light within dielectric materials has persisted over 100 years[1]-[2] and conflicting experiments were reported until recently [3]-[4]. We perform a reversed Fizeau experiment to test the composition law of light speeds in media and the result accords with the extended Lorentz transformation where the light speed $\frac{1}{\sqrt{\varepsilon_0\mu_0}}$ is changed to $\frac{1}{\sqrt{\varepsilon\mu}}$. This is a crucial evidence that Minkowski's formulation $p = n\frac{E}{c}$ should be correct although the momentum is not measured directly.

**Key words:** Abraham-Minkowski controversy; mass-energy equation; Lorentz transformation; graphene; Fizeau experiment; theory of relativity;

**PACS numbers:** 03.30.+p; 42.50.Ct; 41.20.Jb; 04.80.Cc; 03.65.Ta;

1. *Corresponding author, E-mail: zhongyuewang(at)yahoo(dot)com(dot)cn*

## On the Mass-Energy Relation for Photons in Media

This note proves that Einstein's mass-energy relation $e = mc^2$ cannot apply to photons in a medium.

Suppose $\Sigma$ is the density of the photon number in an electromagnetic field. Therefore, the energy density $w$ and mass density $M$ of this field should be $w = \Sigma e$ and $M = \Sigma m$, where $e$ is the energy and $m$ is the equivalent mass of a photon. On the other hand, the energy flow density $\mathbf{S}$ and momentum density $\mathbf{g}$ is defined as $\mathbf{S} = w\mathbf{u}$, where $\mathbf{u}$ is the velocity vector, and $\mathbf{g} = M\mathbf{u}$. Hence,

$$\frac{\mathbf{S}}{\mathbf{g}} = \frac{w}{M} = \frac{e}{m}$$

In classical electrodynamics, the energy flow density of an electromagnetic field in vacuum is

$$\mathbf{S} = \mathbf{E} \times \mathbf{H} = \frac{\mathbf{1}}{\mu_{\mathbf{0}}} \mathbf{E} \times \mathbf{B}$$

where $\mu_0 = 12.566 \times 10^{-7} H/m$, and the momentum density is

$$\mathbf{g} = \varepsilon_0 \mathbf{E} \times \mathbf{B}$$

where $\varepsilon_0 = 8.854 \times 10^{-12} F/m$. Thus, we have

$$\frac{e}{m} = \frac{\mathbf{S}}{\mathbf{g}} = \frac{\mathbf{E} \times \mathbf{B}/\mu_0}{\varepsilon_0 \mathbf{E} \times \mathbf{B}} = \frac{1}{\varepsilon_0 \mu_0} = c^2$$

and this is consistent with the mass-energy relation $e = mc^2$ in the Theory of Relativity.

But in a medium, the dielectric constant is $\varepsilon \neq \varepsilon_0$ and magnetic permeability is $\mu \neq \mu_0$. Here,

$$\mathbf{S} = \mathbf{E} \times \mathbf{H} = \frac{\mathbf{1}}{\mu} \mathbf{E} \times \mathbf{B}, \quad \mathbf{g} = \varepsilon \mathbf{E} \times \mathbf{B}$$

$$\frac{e}{m} = \frac{\mathbf{S}}{\mathbf{g}} = \frac{\mathbf{E} \times \mathbf{B}/\mu}{\varepsilon \mathbf{E} \times \mathbf{B}} = \frac{1}{\varepsilon \mu} \neq c^2$$

Clearly, only in a vacuum where $\varepsilon = \varepsilon_0$ and $\mu = \mu_0$ can this result reduce to Einstein's mass-energy equation.

Zhong Yue Wang
P.O. Box 504, No.144, Tonghe-6 Block
Gongjiang Road, Shanghai 200435, CHINA

## I..... Introduction

What is the momentum of light in dielectric media? According to de Broglie formula of quantum mechanics, $p=\frac{h}{\lambda}=n\frac{hf}{c}=n\frac{E}{c}$ . In special relativity, however, $E=mc^2$ and $p=mV=\frac{E}{c^2}\frac{c}{n}=\frac{E}{cn}=\frac{hf}{nc}$ ( $V$ is the energy flow velocity [20] $\frac{S}{w}=\frac{1}{\sqrt{\varepsilon\mu}}$ equals the phase velocity $\frac{c}{n}=f\lambda$ of a boundaryless medium). Which one goes wrong?

## II..... Resolution

In 2005, scientists of the University of Manchester found the light speed $c$ of the Dirac equation to describe electrons in graphene is replaced by the Fermi velocity $V_F$ whose value is about c/300 [5]. If true, we can use outcomes of the quantum field theory to study condensed matter physics. For instance, the following band-to-band tunneling probability is deduced from W.K.B approximation [6]

$$T_{WKB}\sim\exp(-\frac{\pi E_g^2}{4e\hbar FV_F}) \tag{1}$$

Actually, the exact solution was given by the quantum field theory long ago [7]

$$T\propto\sum_{l=1}^{\infty}\frac{1}{l^2}\exp(-\frac{l\pi\ m_0^2c^3}{e\hbar F}) \tag{2}$$

An energy gap $E_g$ can be treated as twice of the equivalent rest mass of quasiparticles as $E_g=2m_0c^2$ by comparison to Dirac's theory of the antiparticle sea. In addition, the characteristic constant should now be Fermi velocity $V_F$ instead of light speed $c$. Substituing $c\to V_F$ and $E_g=2m_0c^2\to 2m_0V_F^2$ into (2),

$$T\propto\sum_{l=1}^{\infty}\frac{1}{l^2}\exp(-\frac{l\pi E_g^2}{4e\hbar FV_F}) \tag{3}$$

which is just (1) on the condition that $l=1$ [8].

Likewise, we introduce $c/n$ to take the place of $c$ in the Lorentz transformation to investigate electrodynamic and optical phenomena within media [9]~[14]

$$\left.\begin{aligned} x' &= \frac{x-Vt}{\sqrt{1-\frac{V^2}{u^2}}} \\ y' &= y \\ z' &= z \\ t' &= \frac{t-Vx/u^2}{\sqrt{1-\frac{V^2}{u^2}}} \end{aligned}\right\} \quad (u=\frac{1}{\sqrt{\varepsilon\mu}}=\frac{c}{n}) \tag{4}$$

Such idea was also proposed by [15] and [16]. Consequently, the energy and momentum should be

$$E=\frac{m_0}{\sqrt{1-\frac{V^2}{u^2}}}u^2=mu^2 \tag{5}$$

$$p=\frac{m_0}{\sqrt{1-\frac{V^2}{u^2}}}V=mV \tag{6}$$

and $p=\frac{E}{u^2}V=\frac{E}{c^2/n^2}\frac{c}{n}=n\frac{E}{c}=n\frac{hf}{c}$ is consistent with the result of quantum mechanics. Abraham-Minkowski controversy is now resolved. (5) and (6) can be utilized to derive the Cherenkov effect in media.[15]

## III..... Experiment

To test the transformation (4), we modify the design of [13] to re-do Fizeau's experiment underwater for easy of realization. In the original Fizeau convection experiment[17], the fluid inside pipes is water and air outside is still. On the contrary, in our experiment the medium inside is flowing air and external optical path is in rest water. The wavelength of He-Ne laser is $\lambda=632\,nm$ ($f=4.74\times10^{14}\,Hz$). Length of a pipe is $L=4m$ and the speed of air is $V_{air}\approx 9m/s$ .The air pump is placed outdoors to reduce noise and shock. Most of the optical path outside pipes is in two tubes full of water. The picture is as follows (Figure.1),

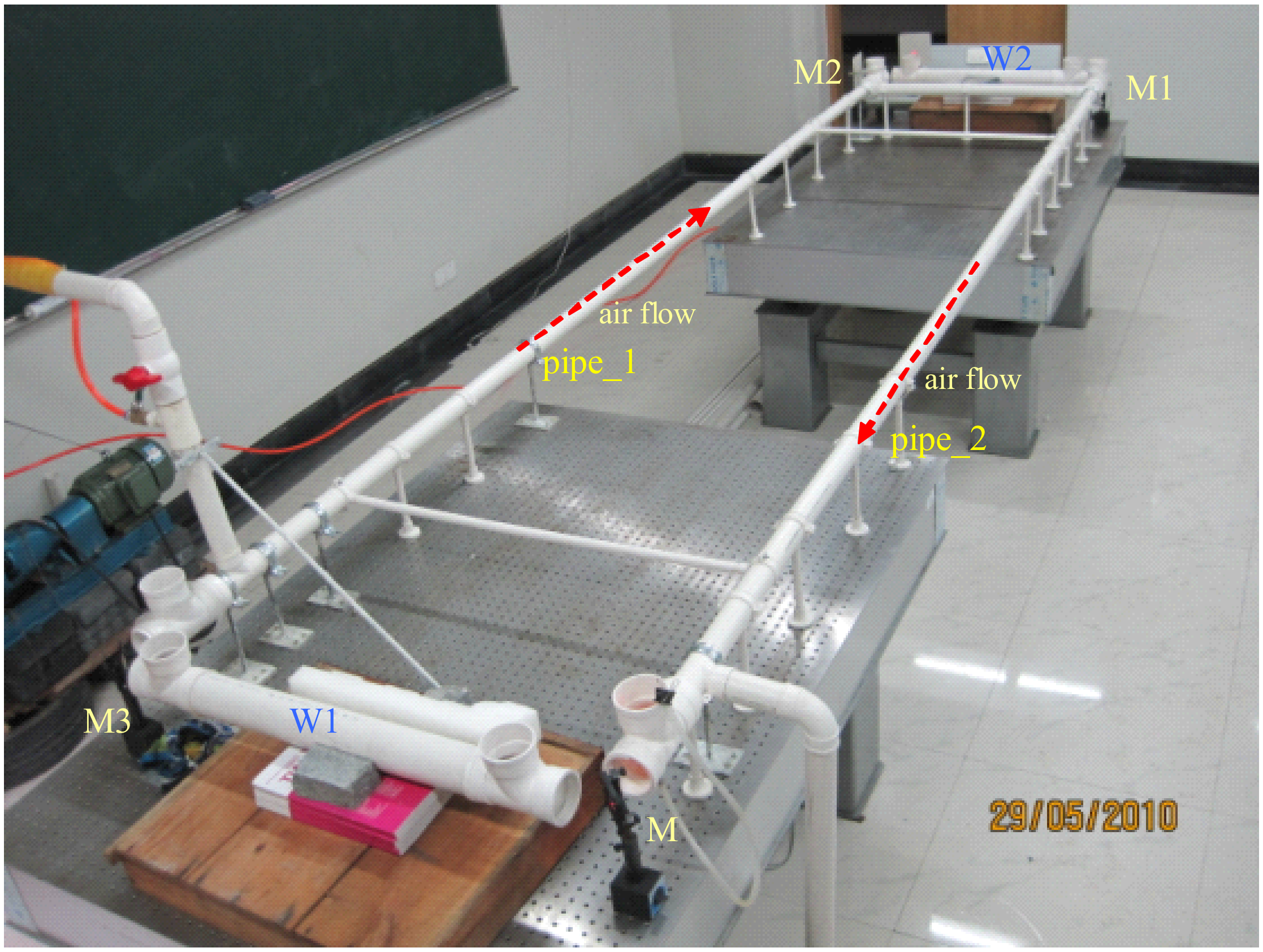

**Figure. 1 Modified Fizeau experiment**

M and M2: reflecting-transmitting mirror; M1 and M3: total reflecting mirror;
W1 and W2: water tube;

Here, we have to change the reflecting mirror $M_2$ in the scheme of [17] to be a half-reflecting and half-transmitting mirror and observe interference fringes behind $M_2$ (Figure.2) in view of the light is too weak to return to the starting point $M$.

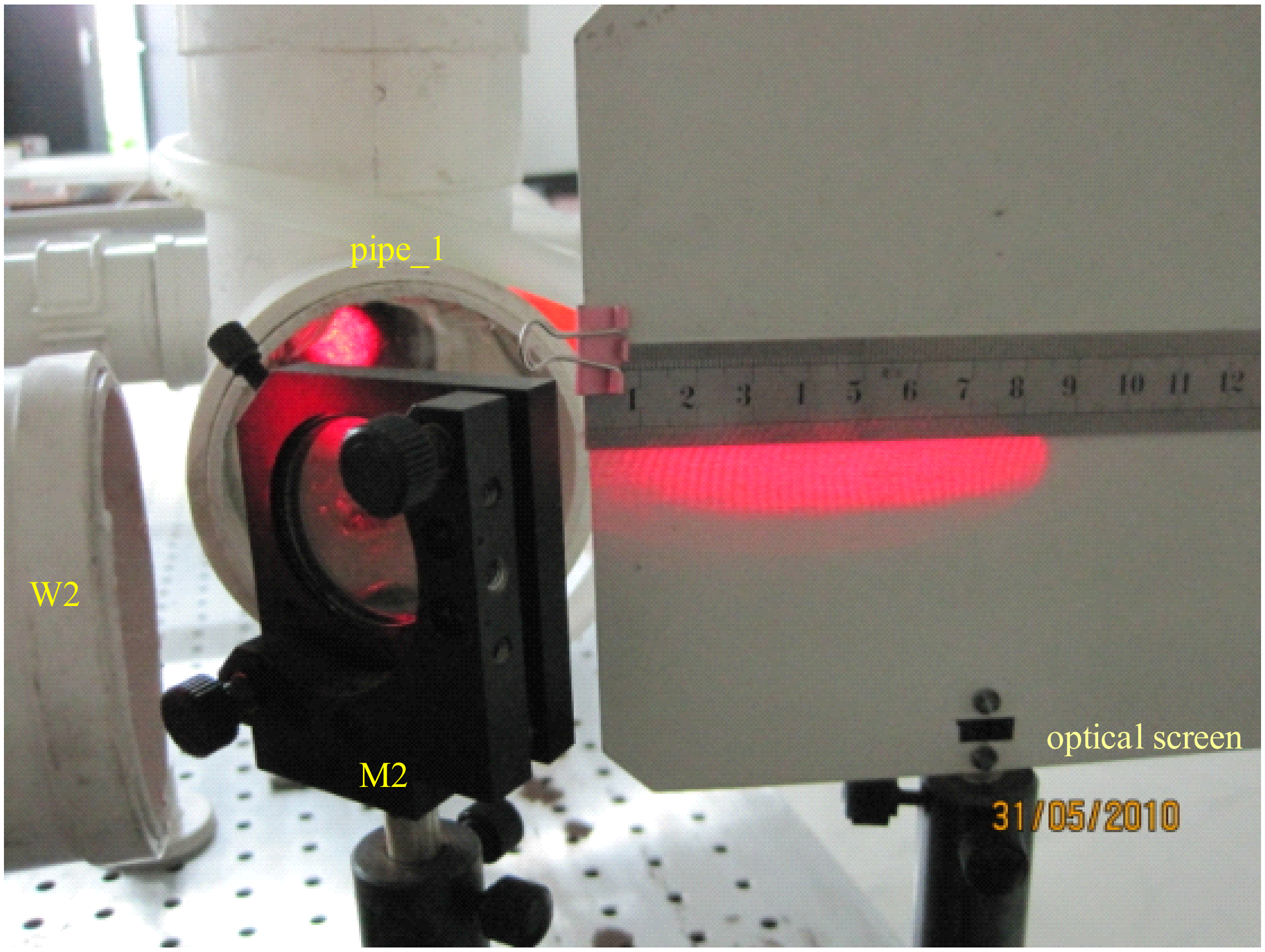

**Figure. 2 Screen**

In fact, this is a Mach-Zehnder interferometer and the total optical path is only half of that in Fizeau's experiment. The time difference between $M \to M_3 \to M_2$ and $M \to M_1 \to M_2$ led by the velocity of air flow is

$$\Delta t = \frac{L}{\frac{c}{n} - kV_{air}} - \frac{L}{\frac{c}{n} + kV_{air}} \approx \frac{2Ln^2 kV_{air}}{c^2}$$

Therefore, the absolute value of the shift of a fringe

$$\Delta N = f\Delta t = \frac{2Ln^2 fkV_{air}}{c^2} \qquad (7)$$

is also half of that in Fizeau's experiment. Fringes corresponding to still and flowing air are in Figs.3 and 4, respectively.

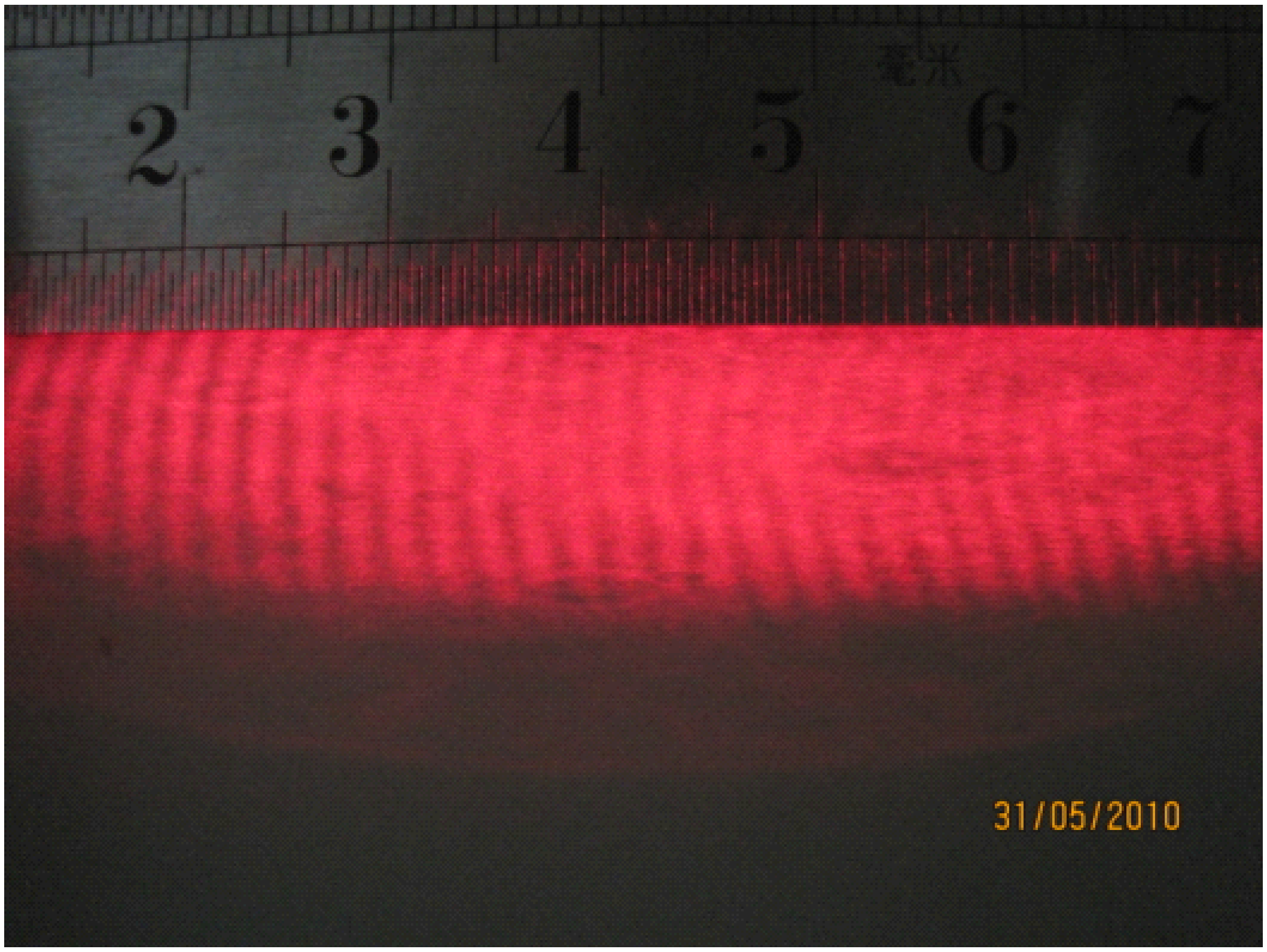

**Figure.3 Air is still**

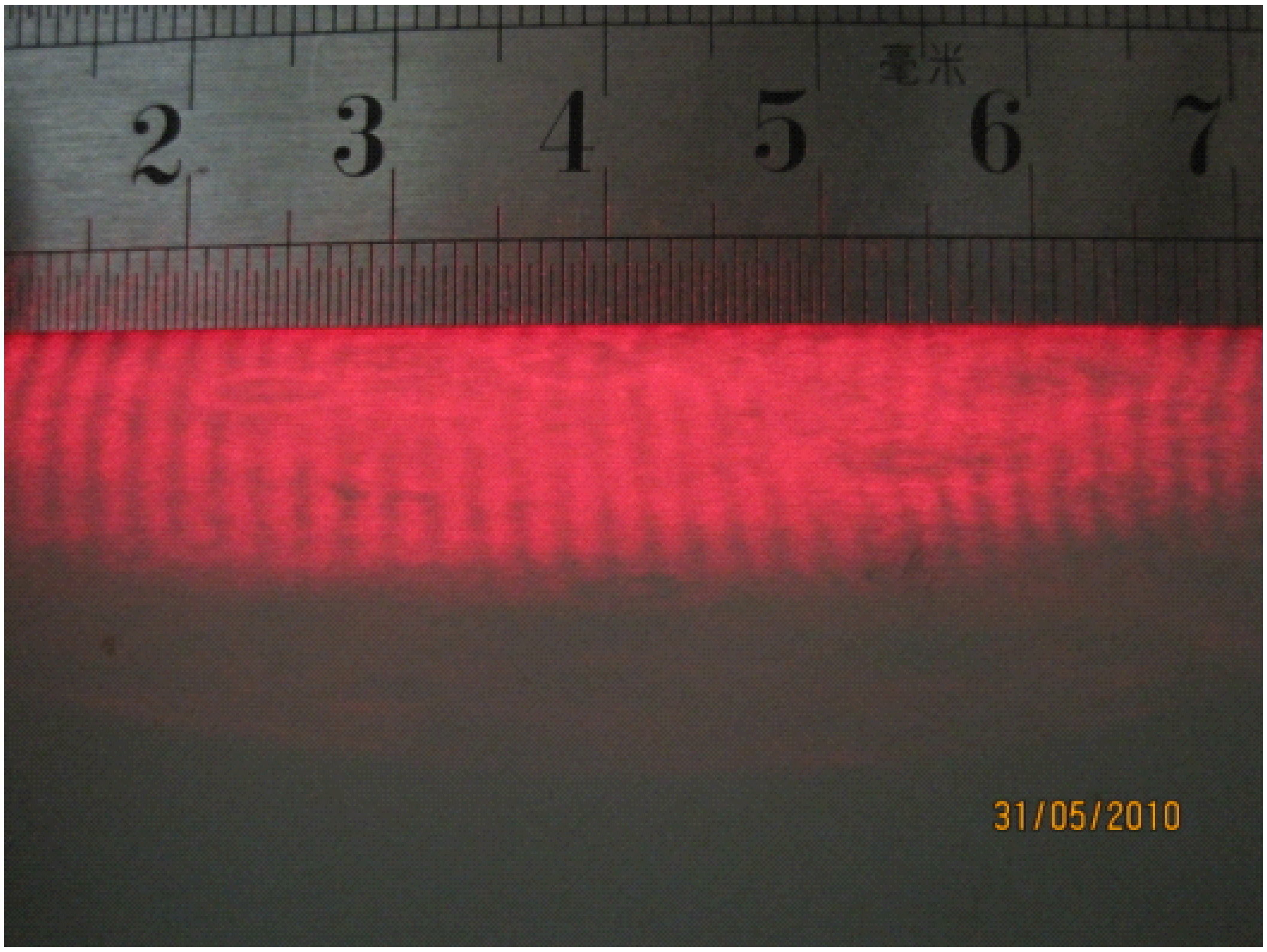

**Figure. 4 Air is flowing**

The average $\Delta N$ is about $1/4$ and not caused by instabilities. Nevertheless, the addition law of velocities from the Lorentz transformation is

$$\frac{\frac{dx}{dt} \pm V}{1 \pm \frac{dx}{dt}\frac{V}{c^2}}$$

In this experiment, $\frac{dx}{dt} = \frac{c}{n}$ (light speed of rest air), $V = V_{air}$

$$\frac{\frac{c}{n} \pm V_{air}}{1 \pm \frac{\frac{c}{n}V_{air}}{c^2}} \approx \frac{c}{n} \pm kV_{air} \qquad (k = 1 - \frac{1}{n^2})$$

Due to the refractive index of air is $n = 1.0003 \approx 1$, $k = 1 - \frac{1}{n^2} = 0.0006 \approx 0$ and the shift $\Delta N \propto k$ is negligible according to (7). It cannot explain the visible shift $\Delta N \approx 0.25$ .

## IV..... Interpretation

Let us apply the following addition law of the new transformation (4) to calculate where $c$ is replaced by the light speed $u$ in water

$$\frac{\frac{dx}{dt} \pm V}{1 \pm \frac{dx}{dt}\frac{V}{u^2}}$$

The refractive index of water is $n' = 1.33$ and light speed is $u = \frac{c}{n'} = \frac{c}{1.33}$,

$$\frac{\frac{c}{n} \pm V_{air}}{1 \pm \frac{\frac{c}{n}V_{air}}{\left(\frac{c}{n'}\right)^2}} \approx \frac{c}{n} \pm kV_{air} \qquad (k = 1 - \frac{n'^2}{n^2})$$

$$\Delta N = \frac{2Ln^2 fkV_{air}}{c^2} = \frac{2L(n^2 - n'^2)fV_{air}}{c^2}$$

Since $n=1.0003\approx 1$ and $n'=1.33$,

$$\Delta N=\frac{2\times 4m\times(1-1.33^2)\times(4.74\times 10^{14}Hz)\times 9m/s}{(3\times 10^8 m/s)^2}\approx -0.3$$

that the negative sign means shifts are in the opposite direction of the Fizeau experiment because the setup is symmetric. In practice, Fizeau's experiment is equivalent to $n=1.33$, $n'=1.0003\approx 1$ and $V_{air}$ gives place to $V_{water}$.

Furthermore,we suggest measuring the gravitational frequency shift $\frac{\Delta f}{f}=\varepsilon_r\frac{gH}{c^2}$ ( $\varepsilon_r=\frac{\varepsilon}{\varepsilon_0}\approx n^2$ ) of microwaves in ultra-high dielectric constant materials on the surface of the earth to test (5) owing to $\Delta E=mgH=h\Delta f$ and $m=\frac{E}{u^2}=n^2\frac{hf}{c^2}\approx\varepsilon_r\frac{hf}{c^2}$ . For example, $\Delta f{\sim}110Hz$ is detectable under conditions of $f=40GHz$, $\varepsilon_r=50{,}000$ and $H=500\,m$ .The refractive index of a $\gamma$ -photon is $n\approx 1$ in the famous Pound-Rebka experiment[18] that is insufficiently to differ $E=mc^2$ and $E=mu^2$ .

## Conclusions

This is to test the extended Lorentz transformation for electromagnetic phenomena in media.It does not contradict all known experiments to confirm the theory of relativity because they were performed in either vacuum ( $n=1$ ) or air,etc ( $n\approx 1$ ). Actually, the greatest contribution of special relativity is neither the Lorentz transformation nor energy-mass equivalence.The soul is that spacetime is not transcendental and depends upon substance. Hence, it is natural to think to exist different spacetime structures whose characteristic constant can be $\frac{1}{\sqrt{\varepsilon_0\mu_0}}=c$, $V_F$, $\frac{1}{\sqrt{\varepsilon\mu}}=\frac{c}{n}$, speeds of neutrinos [8], etc. To any constant $C$, we have two symmetric groups to contain the velocity of motion $V$ from zero to infinity[19]

| $V<C$ | $V>C$ |
|---|---|
| $E=\frac{m_0C^2}{\sqrt{1-\frac{V^2}{C^2}}}$ | $E=\frac{m_0C^2}{\sqrt{\frac{V^2}{C^2}-1}}$ |
| $p=\frac{m_0V}{\sqrt{1-\frac{V^2}{C^2}}}$ | $p=\frac{m_0V}{\sqrt{\frac{V^2}{C^2}-1}}$ |
| $E=\sqrt{p^2C^2+m_0^2C^4}$ | $E=\sqrt{p^2C^2-m_0^2C^4}$ |
| $\frac{E}{p}=\frac{C^2}{V}$ | $\frac{E}{p}=\frac{C^2}{V}$ |

Both lead to $E = pC$ and $m_0 = 0$ at the point of $V = C$. Einstein's theory is only $C = \frac{1}{\sqrt{\varepsilon_0 \mu_0}}$ $= 300{,}000 km/s$ and $V < C$. It cannot exclude the possibility that $C > c$ of undiscovered interactions and $V > c$ even though $C \leq c$ such as the energy flow in the Goos-Hanchen shift [20] and good conductors [21].

.

## Acknowledgments

The author is very grateful to Mr. C.Wang of the College of Information Technology of ZAFU and Dr. X.C.Qin of the East China University of Science and Technology. This experiment is impossible to be carried out without their help. It is also a pleasure to acknowledge Mr. Y.Q.Xu whom is the laboratorian of ZAFU, Mr.J.G.Wu of Laserver Inc. and Mr. J.F.Tong of Zhejiang Brown Sunroofs Co.,Ltd to provide convenience.

## Nomenclature

$f$ = frequency

$\lambda$ = wavelength

$h$ = Planck constant

$c$ = light speed in vacuum

$\varepsilon_0$ = dielectric coefficient of vacuum

$\mu_0$ = magnetic conductivity of vacuum

$u$ = light speed in media

$\varepsilon$ = dielectric coefficient of media

$\mu$ = magnetic conductivity of media

$n = \frac{c}{u}$ = refractive index

$\varepsilon_r = \frac{\varepsilon}{\varepsilon_0}$ = relative dielectric constant

$V_F$ = Fermi velocity

$E_g$ = energy gap

$F$ = electric intensity

$l$ = integer number

$E$ = total energy of a single particle

$m_0$ = rest mass of a single particle

$V$ = velocity of a single particle

$p$ = total momentum of a single particle

$C$ = constant velocities

$g = 9.8 m/s^2$ = acceleration of gravity on the earth of the earth

$H$ = height

**Postscript**: Energy Transport faster than light in Good Conductors

People do not wake up to the fact that the velocity to transfer the electric energy and signal along metal wires is larger than $c$. The theory is based on recent progress of physics that the sample of macroscopic and superluminal energy transport of an electromagnetic field was proposed [1]. Meanwhile, a latest optical experiment confirms the characteristic constant should be $\frac{1}{\sqrt{\varepsilon\mu}}$ within media[2].

## 1. Introduction

A recent paper is to prove [1]

$$\text{wave-particle duality}\left\{\begin{array}{llll} & \text{waveguide} & \text{free space} & \text{surface wave} \\ \textit{field theory} & \frac{S}{w}<c & \frac{S}{w}=c & \frac{S}{w}>c \\ & & & \\ \textit{point mechanics} & V<c & V=c & V>c \end{array}\right\}\text{superluminality}$$

$$\underbrace{\qquad\qquad\qquad\qquad\qquad\qquad}_{\textit{velocity of energy flow } \frac{S}{w}=V \ (\textit{mechanical velocity})}$$

and has nothing to do with other concepts of velocities in the wave theory such as the phase velocity, group velocity and front velocity. Only waves in vacuum are discussed to prevent any argument.

On the other hand, a newest optical experiment [2] implies the parameter $c=\frac{1}{\sqrt{\varepsilon_0\mu_0}}$ of the theory of relativity should be replaced by $u=\frac{1}{\sqrt{\varepsilon\mu}}$ to characterize electromagnetic phenomena within media. Therefore, we apply following equations to restudy Chap.1 of [1].

$$E=\frac{m_0u^2}{\sqrt{1-\frac{V^2}{u^2}}}$$

$$p=\frac{m_0V}{\sqrt{1-\frac{u^2}{u^2}}}$$

$$E^2=p^2u^2+m_0^2u^4 \qquad (1)$$

To compare with the dispersion relation of electromagnetic waves in a closed waveguide filled with the dielectric medium

$$\omega^2=\beta^2u^2+\omega_c^2 \qquad (\beta=k_{//}) \qquad (2)$$

where the cut-off frequency $\omega_c$ is a function depends upon $\frac{1}{\sqrt{\varepsilon\mu}}$ and sectional dimension. Obviously,

rest mass $m_0 u^2 = \hbar\omega_c$

mechanical velocity $V = u\sqrt{1-\frac{\omega_c^2}{\omega^2}} < u$

total energy $$E = \frac{m_0 u^2}{\sqrt{1-\frac{V^2}{u^2}}} = \frac{\hbar\omega_c}{\frac{\omega_c}{\omega}} = \hbar\omega \quad (3)$$

total momentum $$p = \frac{m_0 V}{\sqrt{1-\frac{V^2}{u^2}}} = \hbar\beta \quad (4)$$

ratio of energy to momentum $$\frac{E}{p} = \frac{\omega}{\beta} = \frac{u}{\sqrt{1-\frac{\omega_c^2}{\omega^2}}} > u$$

The velocity of motion $V = u\sqrt{1-\frac{\omega_c^2}{\omega^2}}$ in point mechanics is just the energy flow velocity

$$\frac{P}{U} = \frac{\int\int \overline{S}\, dxdy}{\int\int \overline{w}\, dxdy} = \frac{1}{\sqrt{\varepsilon\mu}}\sqrt{1-\frac{\omega_c^2}{\omega^2}} \quad (5)$$

given by electrodynamics. Besides,

$$\frac{E}{p} = \frac{w/N}{g/N} = \frac{w}{g} \quad (6)$$

$$\frac{\mathbf{S}}{\mathbf{g}} = \frac{\mathbf{E}\times\mathbf{H}}{\varepsilon\mathbf{E}\times\mathbf{B}} = \frac{\mathbf{E}\times\mathbf{B}/\mu}{\varepsilon\mathbf{E}\times\mathbf{B}} \equiv \frac{1}{\varepsilon\mu} = u^2 \quad (7)$$

$$\frac{E}{p} = \frac{\int\int \overline{w}\, dxdy}{\int\int \overline{g}\, dxdy} = \frac{\int\int \overline{w}\, dxdy}{\frac{\int\int \overline{S}\, dxdy}{u^2}} = \frac{u^2}{\frac{P}{U}} = \frac{u^2}{u\sqrt{1-\frac{\omega_c^2}{\omega^2}}} = \frac{u}{\sqrt{1-\frac{\omega_c^2}{\omega^2}}} > u$$

This is in agreement with above ratio between Eq.(3) to Eq.(4)

## 2. Surface electromagnetic waves in dielectric media

Tachyonic equations in a medium

$$E = \frac{m_0 u^2}{\sqrt{\frac{V^2}{u^2}-1}}$$

$$p = \frac{m_0 V}{\sqrt{\frac{V^2}{u^2} - 1}}$$

$$E^2 = p^2 u^2 - m_0^2 u^4 \tag{8}$$

$$\frac{E}{p} = \sqrt{u^2 - \frac{m_0^2 u^4}{p^2}} < u$$

had been utilized to deduce the Cherenkov effect $\cos\theta = \frac{c/n}{V}$ [3]. On account of the dispersion relation of surface electromagnetic waves,

$$\omega^2 = \beta^2 u^2 - \tau^2 u^2 \quad (u = \frac{1}{\sqrt{\varepsilon\mu}},\ \tau^2 > 0)$$

there have

invariant mass $m_0 = \frac{\hbar\tau}{u}$

mechanical velocity $V = u\sqrt{1 + \frac{\tau^2 u^2}{\omega^2}} > u$

total energy $E = \frac{m_0 u^2}{\sqrt{\frac{V^2}{u^2} - 1}} = \hbar\omega$

total momentum $p = \frac{m_0 V}{\sqrt{\frac{V^2}{u^2} - 1}} = \hbar\beta$

$$\frac{E}{p} = \frac{\omega}{\beta} < u$$

Suppose the refractive index of the optically thinner medium in Fig.3 of [1] is $n_2$ ($u_2 = \frac{1}{\sqrt{\varepsilon_2 \mu_2}} = \frac{c}{n_2}, 1 < n_2 < n$), the wave vector should be $k\frac{n}{n_2}$ in the optically denser medium if the value is $k$ in the former. The phase constant is the same one

$$\beta = k\frac{n}{n_2}\sin\theta_i$$

and the dispersion relation of transmitted waves in the optically thinner medium should now be

$$\omega = k u_2 = k\frac{c}{n_2}$$

$$\omega^2 = k^2 \frac{c^2}{n_2^2}$$

$$= \beta^2 \frac{c^2}{n_2^2} - (\frac{n^2}{n_2^2}\sin^2\theta_i - 1)k^2 \frac{c^2}{n_2^2}$$

Owing to $\sin\theta_i > \sin\theta_c = \frac{n_2}{n}$ of total reflection, $(\frac{n^2}{n_2^2}\sin^2\theta_i - 1)k^2 > 0$ which is corresponding to the mechanical equation $E^2 = p^2 u_2^2 - m_0^2 u_2^4$ of tachyons in the optically thinner medium

$$\underbrace{\hbar^2\omega^2}_{E^2} = \underbrace{\hbar^2\beta^2 \frac{c^2}{n_2^2}}_{p^2u_2^2} - \underbrace{(\frac{n^2}{n_2^2}\sin^2\theta_i - 1)\hbar^2 k^2 \frac{c^2}{n_2^2}}_{m_0^2 u_2^4} \qquad (u_2 = \frac{c}{n_2})$$

invariant mass $m_0 = \sqrt{\frac{n^2}{n_2^2}\sin^2\theta_i - 1}\,\frac{\hbar k}{c} n_2$

mechanical velocity $V = u_2 \frac{n}{n_2}\sin\theta_i = c\frac{n}{n_2^2}\sin\theta_i > u_2$

total energy $E = \frac{m_0 u_2^2}{\sqrt{\frac{V^2}{u_2^2} - 1}} = \hbar k u_2 = \hbar\omega$

total momentum $p = \frac{m_0 V}{\sqrt{\frac{V^2}{c^2} - 1}} = \hbar k \frac{n}{n_2}\sin\theta_i = \hbar\beta$

$$\frac{E}{p} = \frac{\hbar\omega}{\hbar\beta} = \frac{\omega}{\beta} = \frac{\omega}{k\frac{n}{n_2}\sin\theta_i} = \frac{u_2}{\frac{n}{n_2}\sin\theta_i} < u_2$$

Moreover, in classical electrodynamics

$$\overline{S_x} = \frac{1}{2}\sqrt{\frac{\varepsilon_2}{\mu_2}}\left|E''_0\right|^2 e^{-2k\sqrt{\frac{n^2}{n_2^2}\sin^2\theta_i - 1}\,z}\ \frac{n}{n_2}\sin\theta_i$$

$$\overline{w} = \overline{w_e} + \overline{w_m}$$

$$= \frac{1}{2}\mathrm{Re}(\frac{1}{2}\varepsilon_2 \mathbf{E}''^* \cdot \mathbf{E}'') + \frac{1}{2}\mathrm{Re}(\frac{1}{2}\mu_2 \mathbf{H}''^* \cdot \mathbf{H}'')$$

$$= \frac{1}{2}\varepsilon_2 \left|E''_0\right|^2 e^{-2k\sqrt{\frac{n^2}{n_2^2}\sin^2\theta_i - 1}\,z}$$

The velocity of the energy density is

$$\frac{\overline{S_x}}{\overline{w}} = \frac{1}{\sqrt{\varepsilon_2\mu_2}}\frac{n}{n_2}\sin\theta_i = c\frac{n}{n_2^2}\sin\theta_i > u_2$$

that can exceed $\frac{1}{\sqrt{\varepsilon_0\mu_0}}$ provided $n_2$ is small enough ( $n_2 < \sqrt{n\sin\theta_i}$ ). In microwave electronics, generally speaking, the optically thinner medium is air whose value is $n_2 \approx 1$ to meet the condition

As to the ratio of energy to momentum,

$$\frac{\overline{w}}{\overline{g}} = \frac{\overline{w}}{\frac{\overline{S}}{u_2^2}} = \frac{u_2^2}{\frac{\overline{S}}{\overline{w}}} = \frac{u_2}{\frac{n}{n_2}\sin\theta_i} < u_2$$

### 3. Electromagnetic field in good conductors

In conductors [4], the wave vector is $k + i\alpha$ ( $i = \sqrt{-1}$ ) and

$$k^2 - \alpha^2 = \omega^2\varepsilon\mu \tag{9}$$

$$k\alpha = \frac{\omega\mu\sigma}{2} \tag{10}$$

The solution is

$$k = \omega\sqrt{\varepsilon\mu}\left\{\frac{1}{2}\left[\sqrt{1+\frac{\sigma^2}{\omega^2\varepsilon^2}}+1\right]\right\}^{1/2} \tag{11}$$

$$\alpha = \omega\sqrt{\varepsilon\mu}\left\{\frac{1}{2}\left[\sqrt{1+\frac{\sigma^2}{\omega^2\varepsilon^2}}-1\right]\right\}^{1/2} \tag{12}$$

and the phase velocity should be

$$V_p = \frac{\omega}{k} = \frac{1}{\sqrt{\varepsilon\mu}}\left\{\frac{1}{2}\left[\sqrt{1+\frac{\delta^2}{\omega^2\varepsilon^2}}+1\right]\right\}^{-1/2} \tag{13}$$

Actually, Eq.(9) is equivalent to Eq.(8) and the above wave theory can be transformed into point mechanics,

invariant mass $m_0 = \frac{\hbar\alpha}{u}$ ( $u = \frac{1}{\sqrt{\varepsilon\mu}}$ )

mechanical velocity $V = u\sqrt{1+\frac{\alpha^2 u^2}{\omega^2}}$ (14)

total energy $E=\dfrac{m_0u^2}{\sqrt{\dfrac{V^2}{u^2}-1}}=\hbar\omega$

total momentum $p=\dfrac{m_0V}{\sqrt{\dfrac{V^2}{u^2}-1}}=\hbar k$

ratio $\dfrac{E}{p}=\dfrac{\hbar\omega}{\hbar k}=\dfrac{\omega}{k}=V_p$

Substituting Eq.(12) into Eq.(14), the velocity of motion in mechanics

$$V=u\sqrt{1+\frac{\alpha^2u^2}{\omega^2}}=\frac{1}{\sqrt{\varepsilon\mu}}\left\{\frac{1}{2}\left[\sqrt{1+\frac{\sigma^2}{\varepsilon^2\omega^2}}+1\right]\right\}^{1/2} \qquad (14')$$

should also be the velocity of energy of the electromagnetic field transferred along the conductor.(Don't take it for the drift velocity of charge carriers which is much more slower). In light of Eq.(13) and Eq.(14') ,

$$V_p.V=\frac{1}{\varepsilon\mu}=u^2 \qquad (15)$$

Since $\dfrac{\sigma}{\omega\varepsilon}>>1$ and $\dfrac{1}{\sqrt{\varepsilon\mu}}$ has the same order of magnitude of $\dfrac{1}{\sqrt{\varepsilon_0\mu_0}}$,

$$V_p\approx\frac{1}{\sqrt{\varepsilon\mu}}\sqrt{\frac{2\omega\varepsilon}{\sigma}}<<c$$

$$V\approx\frac{1}{\sqrt{\varepsilon\mu}}\sqrt{\frac{\sigma}{2\omega\varepsilon}}>>c$$

That is to say, the mechanical velocity $V$ is much faster than $c=\dfrac{1}{\sqrt{\varepsilon_0\mu_0}}$ in good conductors even though $\dfrac{1}{\sqrt{\varepsilon\mu}}<\dfrac{1}{\sqrt{\varepsilon_0\mu_0}}$. For example, $\dfrac{1}{k}\approx\dfrac{1}{\alpha}=\sqrt{\dfrac{2}{\omega\mu\sigma}}$ ( $\mu\approx\mu_0=4\pi\times10^{-7}\mathrm{H}.m$, $\sigma=5.8\times10^{7}\Omega^{-1}\cdot m^{-1}$ ) is $0.85cm$ of **Cu** in the case of $\omega=2\pi\times60Hz$ [5] and the phase velocity should be

$$V_p=\frac{\omega}{k}=2\pi\times60\times0.0085=3.2\,m/s<<c$$

Assume $0.01c^2<\dfrac{1}{\varepsilon\mu}<c^2$ ( $c=\dfrac{1}{\sqrt{\varepsilon_0\mu_0}}=3\times10^8\,m/s$ ), $V=\dfrac{1}{\varepsilon\mu V_p}$ is in the region of $10^6\,c\sim10^8\,c$.

Especially, the velocity approaches infinity with decrease of the frequency and $\sigma\to\infty$. It accords with outcomes of electrodynamics [6] energy flow density $\overline{S}=\frac{k}{2\omega\mu}|E_0|^2 e^{-2\alpha x}$

energy density $\overline{w}=\frac{\varepsilon}{2}|E_0|^2 e^{-2\alpha x}$

product of energy flow velocity $\frac{\overline{S}}{\overline{w}}$ times phase velocity $\frac{\omega}{k}$ $\frac{\overline{S}}{\overline{w}}\cdot V_p=\frac{1}{\varepsilon\mu}=u^2$ to describe TE waves inside a conductor. The velocity of motion Eq.(14') equals the energy velocity $\frac{\overline{S}}{\overline{w}}$ of the field indeed. To transfer the electric energy or information along metallic cables is one kind of mature application technology of power industry and wire communication.It is easy to detect the time difference between the field to traverse the same distance in free space.

## Conclusions

In general, $\frac{1}{\sqrt{\varepsilon\mu}}<\frac{1}{\sqrt{\varepsilon_0\mu_0}}$ and hence other conditions such as boundaries should be introduced to achieve superluminal energy transport for dielectric media.

Comparatively, superluminality in a good conductor is caused by the interaction between the electromagnetic field and regardless of boundary conditions.

The Sommerfeld wire [7] is a good experimental subject because not only surface waves outside [1] but also the field inside the conductor can exceed $c$.